\documentclass[aps,prl,floats, twocolumn,showpacs,superscriptaddress,reprint]{revtex4-1}
\usepackage{graphicx,epsfig}
\usepackage{graphics,dcolumn,bm,epic, eepic,fleqn,float}
\usepackage{amssymb,amsmath,amsfonts,multirow,rotate,color}
\usepackage{soul}
\definecolor{Myorange}{cmyk}{0,0.42,1,0}

\begin{document}

\title{Collective phenomena emerging from the interactions between
  dynamical processes in multiplex networks}

\author{Vincenzo Nicosia}
\affiliation{School of Mathematical Sciences, Queen Mary University of London, 
London E1 4NS, United Kingdom}  
\author{Per Sebastian Skardal}
\affiliation{Department of Mathematics, Trinity College, Hartford, CT 06106, USA}
\author{Alex Arenas}
\affiliation{Department d'Enginyeria Inform\'{a}tica i Matem\'{a}tiques, Universitat Rovira i Virgili, 43007 Tarragona, Spain}
\author{Vito Latora}
\affiliation{School of Mathematical Sciences, Queen Mary University of London, 
London E1 4NS, United Kingdom}  
\affiliation{Dipartimento di Fisica ed Astronomia, Universit\`a di Catania
  and INFN, I-95123 Catania, Italy}

\begin{abstract}
  We introduce a framework to intertwine dynamical processes of
  different nature, each with its own distinct network topology, using
  a multilayer network approach.  As an example of collective
  phenomena emerging from the interactions of multiple dynamical
  processes, we study a model where neural dynamics and nutrient
  transport are bidirectionally coupled in such a way that the
  allocation of the transport process at one layer depends on the
  degree of synchronization at the other layer, and vice versa. We
  show numerically, and we prove analytically, that the multilayer
  coupling induces a spontaneous explosive synchronization and a
  heterogeneous distribution of allocations, otherwise not present in
  the two systems considered separately. Our framework can find
  application to other cases where two or more dynamical processes
  such as synchronization, opinion formation, information diffusion,
  or disease spreading, are interacting with each other.

\end{abstract}


\maketitle

Networks are a powerful way to model and study a wide variety of
complex phenomena~\cite{Strogatz2001Nature,Newman2003SIAM}.
In the recent years, the study of collective dynamical processes on
complex networks has improved our understanding of many complex
systems and shed light on a wide range of
physical, biological and social phenomena including
synchronization~\cite{Dorfler2013PNAS}, disease
spreading~\cite{PastorSatorras2001PRL}, transport~\cite{Noh2004PRL} 
and cascades~\cite{Larremore2011PRL}. Of particular interest in these
works is the interplay between the structure of the network and its
dynamics~\cite{Boccaletti2006PhysRep,Arenas2008PhysRep}.
In fact, the topology of a network has an effect on the dynamical
processes that take place over the network~\cite{Nishikawa2003PRL}, while
some properties of the dynamics can reveal important information on 
the interaction network~\cite{Arenas2006PRL,Rosvall2008PNAS,Nicosia2014}.
Understanding the relations between structure and dynamics can provide
a solid foundation for modeling, predicting, and controlling dynamical
processes in the real world. However, save for a few notable
exceptions~\cite{Buldyrev2010Nature,Granell2013PRL,Czaplicka2016,Peng15}, 
the majority of the studies so far have considered
a single process on a single network, 
ignoring a very important ingredient: often the components of
a complex system interact through two or more dynamics at the same
time, and these dynamics usually depend on each other in highly
non-trivial ways.

In this Letter we propose a general framework for modelling, through a
multiplex network, the {\it coupling of dynamical processes} of the
same type (e.g. the spreading of two coupled diseases) or of
different types (for instance a synchronization dynamics and a
diffusion process).  Moreover, we demonstrate with a specific example
that this coupling mechanism can give rise to the emergence of complex
phenomena generated by the interactions between the different
dynamical processes.  

The natural way to consider $M$ interacting dynamical processes taking
place over a complex system is to use a multiplex network with $M$
layers~\cite{DeDomenico2013,Kivela2014JCN,Battiston2014,Boccaletti2014}.
Each layer contains the same number of nodes, $N$, and there exists a
one-to-one correspondence between nodes in different layers, but the
topology and the very same nature of the connections at each layer may
be different. We then assign a different dynamical process to each
layer. Considering for simplicity the case $M=2$, we assume that the
dynamics of the entire system is governed by the following equations:
\begin{equation}
  \left\{ \begin{array}{c}
   \dot{x}_i = F_{\omega_i} ( {\bf x} , A^{[1]}  )
   \\
   \dot{y}_i = G_{\chi_i} ( {\bf y} , A^{[2]} )
   \\
\end{array}
\right.
\qquad 
i=1,2,\ldots N
\label{eq:gen}
\end{equation}
where ${\bf x} = \{x_1,x_2,\ldots, x_N\} \in \mathbb{R}^N$ and ${\bf
  y} = \{y_1,y_2,\ldots, y_N\}\in \mathbb{R}^N$ denote the states of
the two dynamical processes, while the topologies of the two layers are
encoded in the adjacency matrices $A^{[1]} = \{ a^{[1]}_{ij} \} $ and
$A^{[2]} = \{ a^{[2]}_{ij} \}$ respectively, such that
$a^{[1]}_{ij}=1$ ($a^{[2]}_{ij}=1$) if a link exists between nodes $i$
and $j$ in the first (second) layer, and $a^{[1]}_{ij}=0$
($a^{[2]}_{ij}=0$) otherwise. The dynamical evolution of the two network
processes is ruled respectively by the functions $F_{\omega}$ and
$G_{\chi}$, which depend on the sets of parameters $\omega$ and
$\chi$, so that the state $x_i$ ($y_i$) of node $i$ at the first
(second) layer is a function of the state ${\bf x}$ (${\bf y}$) and of
the topology $A^{[1]}$ ($A^{[2]}$) of the first (second) layer. The
key ingredient that connects the two dynamical processes is provided
by the nature of the correspondence {\it between layers}. In fact, the
parameter $\omega_i$ in function $F_{\omega_i}$ at layer 1 is itself a
function of time which depends on the dynamical state $y_i$ of node
$i$ at layer 2, while the parameter $\chi_i$ at layer 2 depends on the
state $x_i$ of node $i$ at layer 1.  Namely, we have:
\begin{equation}
  \left\{ \begin{array}{c}
   \dot{\omega}_i = f ( \omega_i, y_i) 
   \\
   \dot{\chi}_i = g (\chi_i, x_i)
   \\
\end{array}
\right.
\qquad 
i=1,2,\ldots N
\label{eq:gen2}
\end{equation}
where $f$ and $g$ are two assigned functions.

\begin{figure}[t]
\centering
\epsfig{file =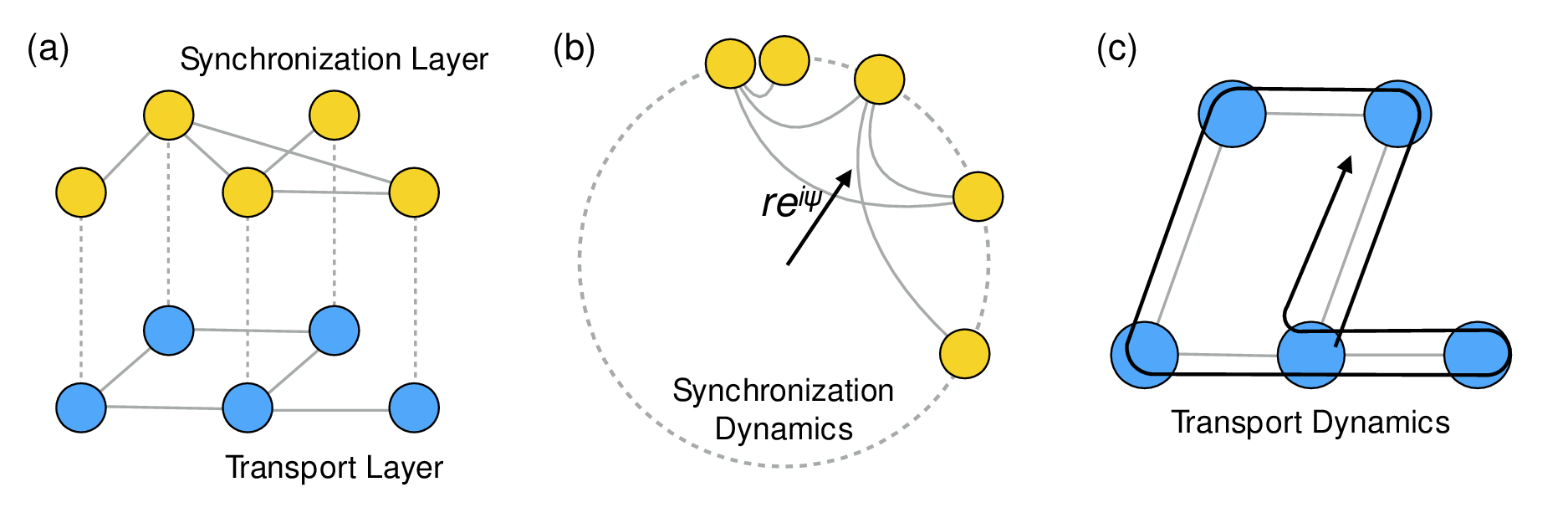,width=0.98\linewidth }
\caption{{\bf Intertwined dynamical processes.} (a) An example of a
  two-layer multiplex of $N=5$ nodes with neural synchronization
  dynamics at layer 1 (top), and transport dynamics at layer 2
  (bottom). (b) The neural activity is described by the Kuramoto model
  in Eq.~(\ref{eq:Kuramoto}), and the degree of synchronization is
  measured by the order parameter $r$. (c) The transport dynamics is
  modelled by biased random walkers moving according to
  Eq.~(\ref{eq:RandomWalk}).  The two dynamical processes are
  bidirectionally coupled, as the natural frequencies of the
  oscillators at layer 1 depend on the distribution of random walkers
  at layer 2 and, at the same time, the random walkers are biased on
  the degree of synchronization of the nodes at layer 1, as described in
  Eqs.~(\ref{eq:OmegaDyn})--(\ref{eq:fDyn}).} \label{fig1}
\end{figure}

As a specific example of this type of coupling, and of the phenomena
that can emerge out of it, we study a toy model of the human
brain. Neural systems depend on the combination of several dynamics,
including blood flow, oxygen exchange, chemical and electrical
interactions among neurons, and remote synchronization of distant
regions~\cite{Varela2001Nature,BullmoreSporns,Deco2013Trends,nico}.
Our multiplex network approach here wants to mimic the interplay
between neural activity and energy transport across brain regions as
illustrated in Figure~\ref{fig1}(a).
Neural activity at the level of brain regions is modelled by the
Kuramoto model~\cite{Kuramoto1984}, such that the state $x_i(t) \in
[0,2\pi)$ of node $i$ at layer 1 corresponds to the phase of
  oscillator $i$ at time $t$, and the first of Eqs.~(\ref{eq:gen})
  reads:
\begin{align}
  \dot{x}_i = \omega_i+ \lambda \sum_{j=1}^N a^{[1]}_{ij}
  \sin(x_j-x_i),
  \label{eq:Kuramoto}
\end{align}
where $\omega_i$ corresponds to the natural frequency of the
oscillator $i$ and $\lambda$ is the {\em coupling strength}.  The
degree of global synchronization in the neural activity is measured by
the Kuramoto order parameter $0 \le r \le 1$ defined by the complex
number $re^{{\rm i}\psi}=\frac{1}{N}\sum_{j=1}^Ne^{{\rm i} x_j}$ which
represents the centroid of all the oscillators on the complex plane.
The second dynamical process, namely energy transport at the second
layer, is modelled by a continuous-time random walk \cite{Lambiotte}.
Specifically, the state $y_i(t)\in [0,1]$ at time $t$ of node $i$ at
the transport layer is equal to the fraction of random walkers at node
$i$ at time $t$, and the second of Eqs.~(\ref{eq:gen}) reads:
\begin{align}
  \dot{y}_i =\frac{1}{\tau_y} \sum_{j=1}^N \left(
  \pi_{ij}-\delta_{ij}\right) y_j = \frac{1}{\tau_y} \sum_{j=1}^N
  \left( \frac{a^{[2]}_{ji} \chi_i^\alpha}{\sum_l a^{[2]}_{jl}
    \chi_l^\alpha} - \delta_{ij} \right) y_j
 \label{eq:RandomWalk} 
\end{align}
where $\pi_{ij}$ is the transition probability from node $j$ to node
$i$, $\tau_y$ is the time scale of the random walker dynamics, and we
have assumed that the random walk is biased on a node property
$\chi_i$, with a tuneable {\em bias exponent}
$\alpha$~\cite{GomezGardenes2008PRE,Sinatra2011,Lambiotte2011PRE,Battiston2016RW}. Notice
that for $\alpha>0$ (resp., $\alpha<0$) the walkers will
preferentially move towards nodes characterised by high (resp., low)
values of $\chi$, while for $\alpha=0$ we recover the standard
unbiased random walk.

To define completely the model, we have to specify how the neural
dynamics and the diffusion of nutrients are coupled, i.e. we need to assign 
the functions $f$ and $g$ in Eqs.~(\ref{eq:gen2}) respectively
relating the frequency $\omega_i$ of the oscillator $i$ at layer 1 to
the available resource $y_i$ at layer 2, and the bias property
$\chi_i$ of the random walkers at layer 2 to the oscillator phase
$x_i$ at layer 1.  First, we assume that the natural frequencies
$\omega_i$, $i=1,2,\ldots,N$, evolve dynamically relaxing to values
proportional to the fraction of random walkers at node $i$ in the
transport layer:
\begin{align}
 \dot{\omega}_i=  \frac { 1 } {\tau_{\omega}} \left(  N y_i(t)-\omega_i   \right),\label{eq:OmegaDyn}
\end{align}
where $\tau_\omega$ gives the timescale for the relaxation. This
choice is motivated by the fact that firing at a higher frequency
usually requires a correspondingly higher amount of energy, in the
form of oxygen and nutrients carried by blood~\cite{Ed2002}.
Next, we assume that the quantities $\chi_i$ evolve according to:
\begin{align}
   \dot{\chi}_i= \frac { 1 } {\tau_{\chi}} \left(  s_i^\text{dyn} - \chi_i \right).\label{eq:fDyn}
\end{align}
where $s_i^{\text{dyn}}$ is the dynamic strength of node $i$, which
measures the local degree of synchronization of oscillator $i$ (degree
to which $i$ is synchronized with its neighbors) and is defined as
$s_i^{\text{dyn}}=r_i\cos(\psi_i-x_i)$ in terms of the local
synchronization order parameter $r_ie^{ {\rm i} \psi_i}=\sum_{j}
a^{[1]}_{ij} e^{ {\rm i} x_j}$. In this way $\chi_i$ relaxes to the
dynamic strength of oscillator $i$ with a timescale $\tau_{\chi}$, and
therefore the random walk described by the transition probabilities in
Eq.~(\ref{eq:RandomWalk}) is biased towards (away from)
strongly-synchronized nodes for positive (negative) values of
$\alpha$. This choice is supported by empirical studies confirming the
existence of correlations between the electrical activity of a brain
area and the hematic inflow in the same area, which is responsible for
the transport of energy to the neurons in the form of oxygen
molecules. In particular, it has been suggested that the high
electrical activity of a brain area is normally followed by an
increase in the blood inflow in the same area~\cite{Malonek1997,
  Sheth2004, Allen2007}.

\begin{figure*}[t]
  \centering
  \includegraphics[width=6in]{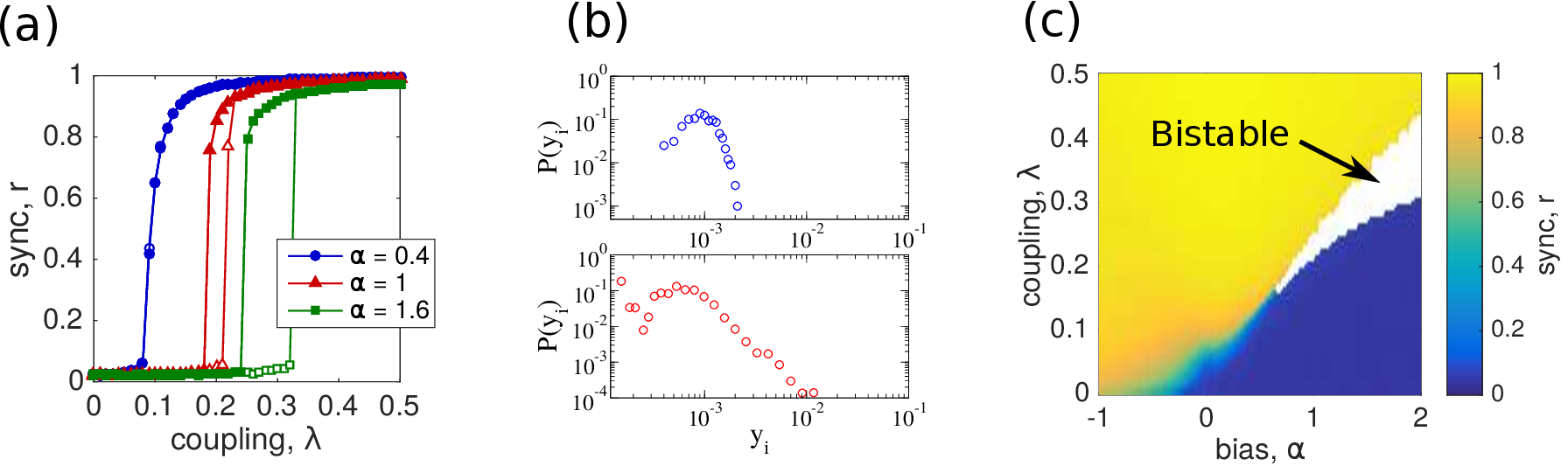}
  \caption{\textbf{Spontaneous explosive synchronization induced by
      the multiplex coupling of the two processes}.  (a) Level of
    synchronization $r$ vs $\lambda$ at layer 1 for bias exponents
    $\alpha=0.4$, $1.0$, and $1.6$ (blue, red, and green,
    respectively).  (b) Distribution $P(y_i)$ of steady-state random
    walker fractions $y_i$ at layer 2 for $\alpha=1.0$, when the
    oscillators at layer 1 are incoherent ($\lambda=0.1$, top, blue)
    and synchronized ($\lambda=0.4$ bottom, red).  (c) Synchronization
    phase diagram showing $r$ as a function of coupling $\lambda$ and
    bias exponent $\alpha$. The bistable region is colored in
    white. Networks are of size $N=1000$ with $\gamma=3$ and $\langle
    k^{[1]}\rangle=\langle k^{[2]}\rangle = 10$.  \label{fig2}
  }
\end{figure*}

Summing up, in the model in Eqs.~(\ref{eq:Kuramoto}-\ref{eq:fDyn}) the
firing rate of a given node $i$ depends on the availability of energy
at $i$ at the transportation layer, and vice versa the abundance of
nutrients at node $i$ depends on the local sychronization of
oscillator $i$ at the neural dynamics layer.  Our model has two
control parameters, $\lambda$ and $\alpha$, that we can change to tune
respectively the coupling between oscillators at layer 1 and the
strength of the bias in the random walk at layer 2.  To illustrate the
effects of intertwining the two dynamical processes, we consider a
multiplex network with $N=1000$ nodes whose synchronization layer is a
scale-free (SF) graph~\cite{Molloy1995} with degree distribution
$P\left(k^{[1]}\right)\propto \left(k^{[1]}\right)^{-\gamma}$ with
$\gamma=3$ above a minimum degree $k_0^{[1]}$, and whose transport
layer is a Erd\H{o}s-R\'{e}nyi (ER) random graph~\cite{Erdos1960} with
link probability $p$.  The average degrees of the two layers are thus
given by $\langle k^{[1]}\rangle=\frac{\gamma-1}{\gamma-2}k_0^{[1]}$
and $\langle k^{[2]}\rangle=p(N-1)$. We choose a SF graph for the
synchronization layer given the prevalence of such topologies in real
neural systems~\cite{Eguiluz05}, and we have considered the limits
$\tau_y,\tau_{\omega},\tau_{\chi}\to0^+$ corresponding to
instantaneous relaxation, meaning the relaxation dynamics of
Eqs.~(\ref{eq:RandomWalk}), (\ref{eq:OmegaDyn}) and (\ref{eq:fDyn}) is
faster compared to the dynamics of the oscillators. These fast
relaxation timescales have been chosen for simplicity, and we note
that the phenomena we present here persists for finite values of
$\tau_y$, $\tau_{\omega}$, and $\tau_{\chi}$, as we show in the
Supplementary Information.

\begin{figure*}[t]
\centering
\includegraphics[width=6in]{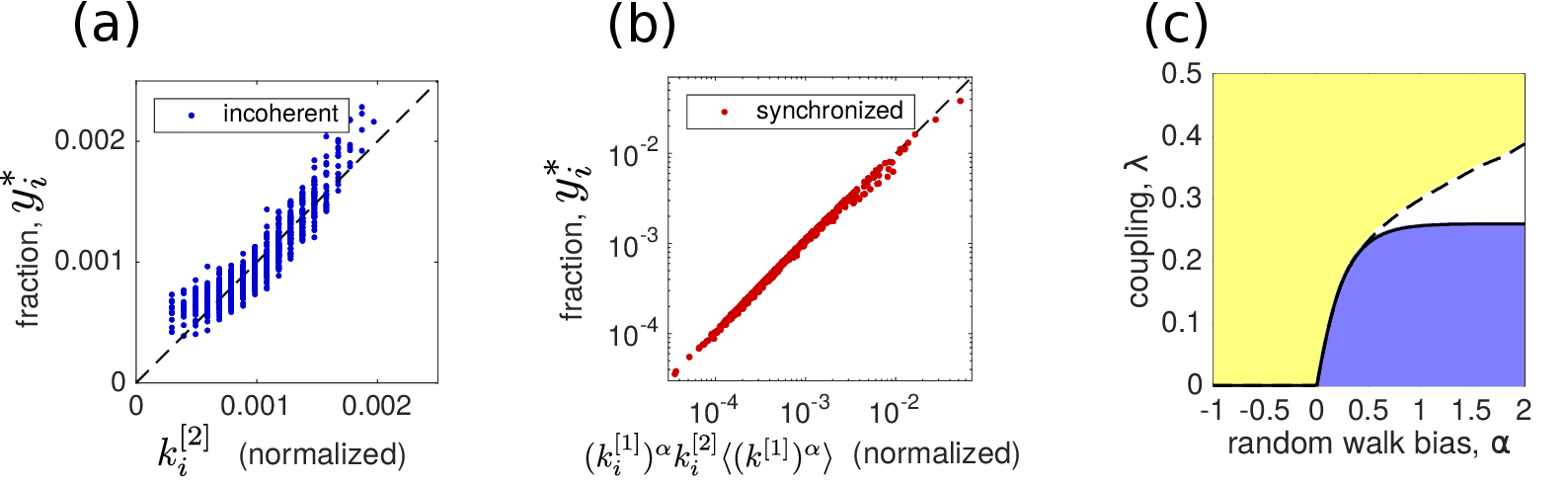}
\caption{{\bf Analytical approach to explain the observed
      collective phenomena.}  Fraction $y_i$ of random walkers at
  node $i$ vs $k_i^{[2]}$ for an incoherent state (a), and vs.
  $(k_i^{[1]})^\alpha k_i^{[2]}\langle (k^{[1]})^\alpha\rangle$ for a
  synchronized state.  (c) Analytically obtained synchronization phase
  diagram showing $r$ as a function of $\lambda$ and $\alpha$.
  Networks are of size $N=1000$ with $\gamma=3$ and $\langle
  k^{[1]}\rangle=\langle k^{[2]}\rangle= 10$ as in the numerical
  simulations shown in Figure~\ref{fig2}.} \label{fig3}
\end{figure*}

We simulated the model on networks with $\langle 
k^{[1]}\rangle=\langle k^{[2]}\rangle=10$, by adiabatically increasing
and then decreasing the coupling strength $\lambda$ at fixed 
values of the bias parameter $\alpha$.
In Figure~\ref{fig2}(a) we report the synchronization profiles $r$ vs
$\lambda$ for $\alpha=0.4$, $1.0$, and $1.6$ (blue circles, red
triangles, and green squares respectively) at layer 1. Notice that for
$\alpha=0.4$ we have the typical continuous phase transition of the
Kuramoto model. Conversely, for $\alpha=1.0$ and $1.6$ we observe the
emergence of a switch-like {\em explosive
  synchronization}~\cite{GomezGardenes2011PRL} and a bistability in
the form of a hysteresis loop (in the forward and backward branches of
the profiles).
In Figure~\ref{fig2}(b) we focus on layer 2, and we plot the
distribution $P(y_i)$ of the steady-state random walker occupation
probabilities $y_i$ for $\alpha=1$, corresponding respectively to
$\lambda=0.1$ when the system at layer 1 is in an incoherent state
(top, blue), and to $\lambda=0.4$ when the system at layer 2 is
synchronized (bottom, red).  While the values of $y_i$ are relatively
homogeneous in the incoherent state and span less than a decade, in
the synchronized state the distribution is heterogeneous and spanning
several decades.  Finally, in Figure~\ref{fig2}(c) we explore the
$(\alpha,\lambda)$ parameter space in more detail, plotting the value
of $r$ at layer 1 as a function of the two control parameters of the
model.  The bistable region which emerges at $\alpha\approx0.7$ and
widens by increasing $\alpha$ is reported in white. We note that this
behavior persists under a wide range of network topologies, provided
that the synchronization layer is sufficiently heterogeneous, as shown
in the Supplementary Information.

Our results indicate that the intertwined nature of diffusion process and
synchronization dynamics gives rise to the emergence of phenomena not
present if the two dynamics were not coupled. Namely, in the transport
layer, we observe a transition from a homogeneous to a heterogeneous
distribution of the random walkers throughout the network, according
to whether the oscillators at the other layer are incoherent or
synchronized.  Concurrently, when the random walkers are biased
sufficiently strongly towards regions that are more synchronized, the
heterogenous distribution of random walkers fosters the emergence of
switch-like {\em explosive
  synchronization}~\cite{GomezGardenes2011PRL} in the neural dynamics
layer. The resulting phase diagram exhibits three phases (incoherent,
bistable, and synchronised) and a tricritical point.  It is noticeable
that explosive synchronization appears naturally in our model due to
the intertwined dynamics of the two processes, and it does not require
ad hoc externally imposed correlations between the oscillator
frequencies and the topology of the interaction network, as those
necessary instead in a single layer network with a single
dynamics~\cite{GomezGardenes2011PRL,Leyva2012NatureSR}.

We now demonstrate that, despite the inherent intricacy of the model,
its dynamical behaviour can be understood analytically. In particular,
we search for conditions such that random walker probabilities and
local order parameters are in a stationary state, $y_i= y_i^*$ and
$r_i=r_i^*$. A steady-state analysis can then be carried out for both
the transport and synchronization dynamics, which we detail in the
Supplementary Information. In particular, we find that the fraction of
random walkers $y_i^*$ depends on whether the synchronization dynamics
is incoherent or synchronized, namely:
\begin{align}
 y_i^*\propto\left\{
\begin{array}{cl}
k_i^{[2]} & \text{ if }r\approx0\\
\left(k_i^{[1]}\right)^\alpha k_i^{[2]} \langle \left(k^{[1]}\right)^\alpha\rangle & \text{ if }r\approx1.
\end{array}\right.\label{eq:Theory02}
\end{align}
Also, the global order parameter $r$ can be written implicitly in terms of the collective frequency $\Omega=\langle\omega\rangle$ and the joint degree-frequency distribution $P(k,\omega)$:
\begin{align}
r &= \frac{1}{\langle k^{[1]}\rangle}\iint\displaylimits_{|\omega-\Omega|\le\lambda rk^{[1]}}\!\!\!\!\!\!P(k^{[1]},\omega)k^{[1]}\sqrt{1-\left(\frac{\omega-\Omega}{\lambda rk^{[1]}}\right)^2}d\omega dk^{[1]},\label{eq:Theory03}
\end{align}
which depends on the topologies of both layers since $\omega_i=Ny_i$
in the steady-state. Figure~\ref{fig3} shows that our analytical
results are in good agreement with the numerical simulations. In
Figures~\ref{fig3}(a) and \ref{fig3}(b) we plot the observed fraction
$y_i$ of random walkers at the steady state vs the predictions of
Eq.~(\ref{eq:Theory02}), respectively for the incoherent and
synchronized state. Dashed black lines are plotted to guide the eye.
In Figure~\ref{fig3}(c) we report the synchronization phase digram
obtained from Eq.~(\ref{eq:Theory03}). A comparison with the phase
diagram in Figure \ref{fig2}c indicates that our theory is able to
accurately reproduce the collective phenomena emerging from the
interactions of the two dynamical processes that we have observed in
our numerical simulations.

The specific example of intertwined synchronization and
transport dynamics studied here shows that interesting collecting
behaviors can appear when we couple two dynamical processes taking place 
on the same set of nodes. Namely, we have found that the
distribution of random walkers in the transport network changes from
homogeneous to heterogeneous according to whether the synchronization
dynamics is incoherent or synchronized, and this result is unexpected
since for the topology of the transport network  
we have on purpose chosen a homogenous graph. 
At the same time,
the heterogeneous distribution of walkers is responsible for the emergence
of explosive synchronization, and the appearance 
of a bistable phase and of a tricritical point in the 
neural network layer. Importantly, here, explosive synchronization 
spontaneously emerges from the interactions of the two dynamical 
processes, without any externally imposed assumptions, necessary instead 
in networks where the Kuramoto model is not coupled to other dynamical
systems~\cite{GomezGardenes2011PRL,Leyva2012NatureSR}.

The switch-like transition we have found in our model closely mirrors
that displayed by the human brain~\cite{Deco2013Trends}, which has the
ability to very quickly switch between resting state activity
(i.e., the background activity
of a brain when no particular conscious task is performed) and complex
intellectual/motor tasks~\cite{Raichele2001}, and thus
requires a fast and flexible mechanism to induce a sudden and
massive synchronization.
 The choice of this specific model was motivated by the
important role that synchronization and transport play in a wide range
of natural and man-made
systems~\cite{Restrepo2005PRE,Motter2013NaturePhys,Skardal2015SciAdv,Gleich2015SIAM} and by the various
bistabilities empirically observed in physics, biology and
neuroscience~\cite{Martens2013PNAS,Brandman2005,Deco2013Trends}.
To date, several studies have investigated how a single type of
dynamics evolves on a multilayer
network~\cite{Sahneh2014PRE,Lee2014PRE,Posfai2016PRE,Burkholz2016PhysicaD}.
However, the framework we have proposed here, based on the use of
multiplex networks to mutually couple dynamics of different nature, is
very general and versatile. We believe that further studies of other
intertwined dynamical processes will uncover other novel phenomena
induced by multiplex coupling, and will eventually result in a more
thorough understanding of the relation between the structure and the
dynamics of multidimensional complex systems.

\begin{acknowledgments}
A.A. and S.S. acknowledge support from MULTIPLEX, grant number 317532 of
the European Commission. V.L. and V.N. acknowledge support from
LASAGNE, grant number 318132 funded by the European Commission. 
V.L. acknowledge support from the EPSRC projects GALE, EP/K020633/1, 
and EP/N013492/1.   
A.A. acknowledges Spanish Ministerio de Economiia y Competitividad,
grant number FIS2015-71582-C2-1, ICREA Academia and the James
S. McDonnell Foundation.
\end{acknowledgments}

\end{document}